\documentclass[twocolumn]{jpsj3}
\usepackage{txfonts}
\usepackage{bm}
\usepackage{graphicx}
\usepackage{braket}
\usepackage{here}
\usepackage{color}
\usepackage{breqn}

\makeatletter
\let\cat@comma@active\@empty
\makeatother

\begin{document}
\title{Material Optimization of Potential High-$T_{\text{c}}$ Superconducting Single-layer Cuprates}

\author{Shingo Teranish$^1$\thanks{shingoteranishi@gmail.com}, 
 Kazutaka Nishiguchi$^2$, and Koichi Kusakabe$^1$}
\inst{
$^1$Department of Materials Engineering Science, Graduate School of Engineering Science, Osaka University, 1-3 Machikaneyama-cho, Toyonaka, Osaka 560-8531, Japan\\
$^2$Department of Physics, Graduate School of Science, Osaka University, 1-3 Machikaneyama-cho, Toyonaka, Osaka 560-8531, Japan}

\abst{
We investigated the material parameters of several single-layer cuprates, including those with fluorinated buffer layers, with the aim of identifying possible high-temperature superconductors. To evaluate the material parameters, we use the Wannierization techniques and the constrained random phase approximation. The obtained single-band Hubbard models are studied using the fluctuation-exchange approximation. Comparison among several cuprates reveals unknown high-$T_{\text{c}}$ superconductors. In, Ga, Al, and Cd compounds in particular show the potential to exhibit higher-$T_{\text{c}}$ superconductivity than Hg1201.
}

\maketitle

\section{Introduction}
Among single-layer cuprates possessing one single CuO$_2$ layer in a unit cell,  HgBa$_2$CuO$_{4+\delta}$ holds a world record critical temperature for superconductivity ($T_{\text{c}}$) for more than 20 years\cite{Hg1201-first}. Several studies have attempted to reveal the factors determining $T_{\text{c}}$ in order to understand material dependence of $T_{\text{c}}$ and to find higher-$T_{\text{c}}$ cuprates. From a theoretical aspect, downfolding into a Hubbard model using the maximally-localized Wannier functions is one of the established methods in which the material parameters of the effective model such as the transfer integral of single-electron hopping
$t$ and the on-site Coulomb interaction $U$ are estimated as a series of the decisive factors for cuprates. For example, an indicator of the shape of the Fermi surface ($r = |t'|+|t''|/|t|$) and the energy difference between the d$_{x^2 - y^2}$ and d$_{3z^2-r^2}$ Wannier orbitals ($\Delta E$)\cite{Sakakibara2010,sakakibara2012,Sakakibara2014} are known as important factors in superconductors mediated by spin fluctuations. The general trend is that the superconductivity becomes stronger when the shape of the Fermi surface is less rounded (smaller $r$) and the energy difference between the d$_{x^2 - y^2}$ and d$_{3z^2-r^2}$ Wannier orbitals is larger (larger $\Delta E$).

In our previous research, we found that these two factors alone cannot fully explain the different $T_{\text{c}}$'s between the two single-layer cuprates HgBa$_2$CuO$_4$
(Hg1201) and TlBa$_2$CuO$_5$ (Tl1201)\cite{cRPAteranishi}. 
Actually, the material dependence of $U$ enables us to explain the experimentally measured values of $T_{\text{c}}$. Here, we summarize the previous research. Although $T_{\text{c}}$ of Hg1201 is higher than that of Tl1201, both $r$ and $\Delta E$ in Hg1201 are worse than those in Tl1201 according to the Sakakibara criterion\cite{Sakakibara2010,sakakibara2012,Sakakibara2014}.
Therefore, there has to be another decisive factor for $T_{\text{c}}$. Really, our thoretical calculation shows that $U$ of Hg1201 is relatively larger than that of Tl1201. As the result, the superconducting strength of the Hg compound is stronger than the Tl one. 
We therefore conclude that $T_{\text{c}}$ in Hg1201 is higher than Tl1201, owing to the stronger $U$. 

The result suggests that the strength of $U$ can be modified by changing or introducing the substitutional doping in the buffer layer. Namely, the components and dopants in the buffer layer can control not only the carrier concentration in the CuO$_2$ layer, but also relevant material parameters that affect the superconductivity. In the previous study, we concluded that, by oxygen control in the Tl layer the strength of the Coulomb interaction may become stronger, thereby enhancing $T_{\text{c}}$.

Motivated by previous research showing the importance of the components and dopants in the buffer layer, we analyze several cuprates including hypothetical compounds with chemical elements in groups 12 and 13, such as Cd, Zn, In, Ga and Al. Here we adopt fluorination of the buffer layers for the Tl, In, Ga, and Al compounds to adjust the doping rate in the CuO$_2$ layer to be the half-filling.

The electronic structures are investigated by the first-principles calculations based on the density functional theory (DFT) with the generalized gradient approximation (GGA). Then the maximally localized Wannier function is employed with the Wannierization techniques\cite{wannier_ent,wannier_loc,wannier90} to evaluate the hopping parameters for the 3d$_{x^2-y^2}$ band ($t,t',t''$), and the constrained Random Phase Approximation (constrained-RPA)\cite{cRPA,cRPA_cuprates_Aryasetiawan} is applied to estimate the strength of the screened on-site Coulomb interaction  ($U_{\mathrm{screened}}$) and the bare on-site Coulomb interaction ($U_{\mathrm{bare}}$) for the cuprates. Additionally, we calculate the eigenvalue of the linearized Eliashberg equation of the single-band Hubbard model on a square lattice by applying the fluctuation-exchange (FLEX) approximation\cite{FLEX_1,FLEX_2}. We also discuss the structural stability of materials by evaluating the total and formation energies.

\section{Methods}
\begin{figure}[ht]
    \includegraphics[keepaspectratio,width=80mm]{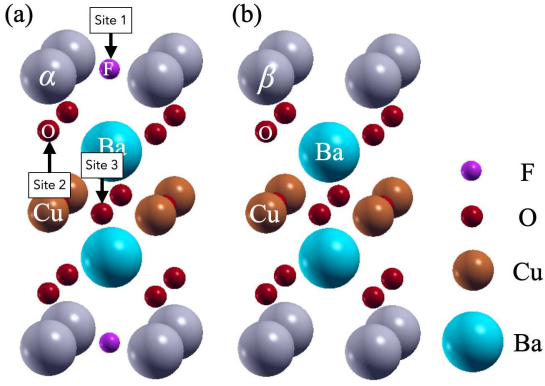}
\caption{Atomic structures of (a)$\alpha$Ba$_2$CuO$_4$F and (b)$\beta$Ba$_2$CuO$_4$ with $\alpha$=Tl,In,Ga,Al and $\beta$=Hg,Cd,Zn. In TlBa$_2$CuO$_5$, F of $\alpha$Ba$_2$CuO$_4$F is replaced by O.\label{fig:str}}
\end{figure}
\begin{figure}[ht]
    \includegraphics[keepaspectratio,width=90mm]{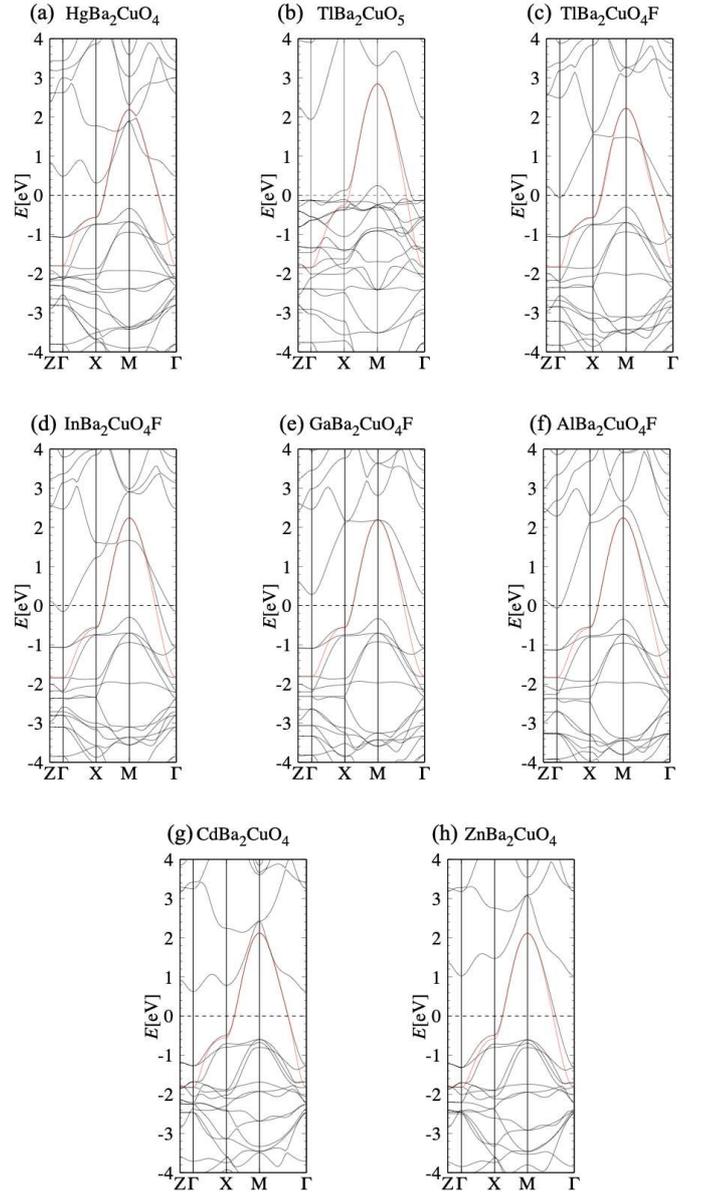}
        \caption{Band structures of (a)HgBa$_2$CuO$_4$, (b)TlBa$_2$CuO$_5$, (c) TlBa$_2$CuO$_4$F, (d) InBa$_2$CuO$_4$F, (e) GaBa$_2$CuO$_4$F, (f) AlBa$_2$CuO$_4$F, (g) CdBa$_2$CuO$_4$ and (h) ZnBa$_2$CuO$_4$.  Wannier-interpolated bands are plotted with a red line. Here, the Fermi level is set to 0.\label{fig:band_eight}}
\end{figure}
In this research, we analyze $\alpha$Ba$_2$CuO$_4$F ($\alpha$1201F) with $\alpha$=Tl, In, Ga, Al and $\beta$Ba$_2$CuO$_4$ ($\beta$1201) with $\beta$=Hg, Cd, Zn. We also included the results of TlBa$_2$CuO$_5$ for comparison. The crystal structures of these systems are shown in Fig.~\ref{fig:str}.  
Fluorination on the buffer layer has already been achieved \cite{Hg1201_F_Tc}. From a viewpoint of ionization of atoms in a crystal, two (one) holes are usually doped per inserted oxygen (fluorine) atom.

Now, let us discuss filling factors of these cuprates. A nominal filling factor in the CuO$_2$ planes can be obtained by the rules for valence in ionized elements and the total charge neutrality.
Assuming that the group 12 elements possess divalence,  $\beta^{+2}$($\beta$=Hg,Cd,Zn), the group 13 elements have trivalence, $\alpha^{+3}$($\alpha$=Tl,In,Ga,Al), and Barium, oxygen, and fluorine atoms are ionized as Ba$^{+2}$, O$^{-2}$, and F$^{-1}$, the formal valences of Cu for these compounds are +2 (half-filling) for $\alpha$Ba$_2$CuO$_4$F and $\beta$Ba$_2$CuO$_4$. For TlBa$_2$CuO$_5$, the formal valences of Cu becomes +3.

The electronic structures of these cuprates can be obtained from the first-principles calculations with Quantum ESPRESSO\cite{QE-2009,QE-2017}. Then, we evaluate the hopping parameters among the 3d$_{x^2-y^2}$ using the Wannierization techniques\cite{wannier_ent,wannier_loc,wannier90}. In this work, we use the RESPACK code to perform the constrained-RPA calculation\cite{respack_1,respack_2,respack_3,respack_4,respack_5}.

In the calculations, we adopt the norm-conserving pseudopotentials and the Perdew-Burke-Ernzerhof functional, and the wave function and charge density are expanded in plane waves with the cutoff energies of (1360,5440) [eV] and the cutoff energy for the polarization functions is 136 [eV]. We use the $8\times8\times8~k-$point mesh in the first Brillouin zone and take into account 100 bands in all calculations. The unit cell volume and atomic positions of each compound are optimized in the simulation, where the pressure is controlled using the criterion for each diagonal element of the stress tensor being less than 0.5 [kbar]. The internal atomic structures are optimized based on the criterion that the summation of the absolute values of force-vector elements becomes smaller than $2.6\times10^{-7}$ [eV/\AA].


To consider the strength of superconductivity, we investigate the obtained Hubbard models within the FLEX approximation and solve the linearized Eliashberg equation. 
A single-band Hubbard Hamiltonian is represented as
\begin{align}
H=\sum_{ij\sigma}\left( t_{ij}c^\dag_{i\sigma}c_{j\sigma}+H.c.\right)+U\sum_in_{i\uparrow}n_{i\downarrow},
\end{align}
where $c^\dag_{i\sigma}$ ($c_{i\sigma}$) represents the creation (annihilation) operator of electrons at site $i$ on a square lattice with spin $\sigma$. $n_{i\sigma}$ represents the particle-number operator at site $i$ and spin $\sigma$. The energy dispersion $\xi_{\bm k}$ of the tight-binding model is
\begin{dmath}
\xi_{\bm k} = 2t(\cos{k_x}+\cos{k_y}) + 4t'\cos{k_x}\cos{k_y} +2t''(\cos{2k_x}+\cos{2k_y})-\mu,
\end{dmath}
where $\mu$ represents the chemical potential and $t$,$ t'$, and $t''$ represent the nearest-, second-, and third-neighbor hopping, respectively.

The interacting Green's function $G(k) $ is obtained as
\begin{equation}
G(k) = \left[G_0(k)^{-1}- \Sigma(k)\right]^{-1},
\label{dyson}
\end{equation}
where $G_0(k)$
is the non-interacting Green's function,
\begin{equation}
G_0(k) = \frac{1}{i\varepsilon_n-\xi_{\bm k}}.
\end{equation}
Here $k=(\bm{k},\mathrm{i}\varepsilon_{n})$, and  $\varepsilon_{n}=(2n-1)\pi k_{\mathrm B}T$ denotes the the Matsubara frequencies for for fermions.
The spin susceptibility within the FLEX approximation obtained as 
\begin{equation}
\chi_{\mathrm{spin}}(q)=\frac{\chi_0(q)}{1-U\chi_0(q)},
\end{equation}
where
\begin{equation}
\chi_0(q) \equiv -\frac{ k_{\mathrm B}T}{N}
\sum_{k}G_0(k
)G_0(k+q)
\end{equation}
is the irreducible susceptibility. 
Here, $q = (\bm{q},i\omega_m)$, and $\omega_m=2\pi mk_{\mathrm B}T$ denotes the the Matsubara frequencies for bosons. $N$ is the number of sites, and $T$ is the temperature. 
The self-energy is obtained as 
\begin{equation}
\Sigma(k)=\frac{ k_{\mathrm B}T}{N}\sum_{q}V^{\Sigma}(q)G(k-q),
\label{self}
\end{equation}
with
\begin{dmath}
\nonumber
V^{\Sigma}(q)=U^2\left[\frac{3}{2}\left(\frac{\chi_0(q)}{1-U\chi_0(q)}\right)\\+\frac{1}{2}\left(\frac{\chi_0(q)}{1+U\chi_0(q)}\right)-\chi_0(q)\right],
\end{dmath}
within the FLEX approximation.

In FLEX approximation, the interacting Green's function $G(k) $ is determined self-consistently. In the self-consistent calculations, the filling is always fixed to be a given electron concentration by adjusting the chemical potential $\mu$.

The linearized Eliashberg equation for the superconducting gap function $\Delta(k)$ is given as
\begin{equation}
\lambda\Delta(k)=-\frac{k_BT}{N}\sum_{k'} V^s(k-k')G(k')G(-k')\Delta(k').
\end{equation}

Here, $\lambda$ is the eigenvalue, where $\lambda=1$ corresponds to $T=T_{\text{c}}$; thus, $\lambda$ serves as a measure of the strength of superconductivity. $V^s$ is the effective interaction for a spin-singlet pairing and is represented as 
\begin{dmath}
V^s(q)=U^2\left[\frac{3}{2}\left(\frac{\chi_0(q)}{1-U\chi_0(q)}\right)-\frac{1}{2}\left(\frac{\chi_0(q)}{1+U\chi_0(q)}\right)\right]+U.
\end{dmath}

In our calculations, we take $N=128^2$ sites with 2,048 Matsubara frequencies, and set $k_{\mathrm B}T=0.03$~eV. We consider filling factors of $n=0.85~(15\%$~hole~doping) because $T_{\text{c}}$ takes maximum around 15\% doping.

Determination of the band parameters ($t$) is done using the mother compounds which are basically at the half filling of the CuO$_2$ plane except for TlBa$_2$CuO$_5$. The rate of change of hopping parameters by small rates of the substitutional dopants in the buffer layer is assumed to be the same across materials so the superconducting strength is then counted by the eigenvalue $\lambda$ at the 15\%-doping in the Hubbard model. This method gives us a reasonable estimation of the superconductivity because the spin-fluctuation mechanism can explain the material dependence except for the under-doped region of the phase diagram. 


\section{Results}
\subsection{Parameter evaluation for $\Delta E$, $r$ and $U$}
\begin{table*}[ht]
  \caption{The Screened on-site Coulomb interaction ($U_{\mathrm{screened}}$) and the bare on-site Coulomb interaction ($U_{\mathrm{bare}}$), hopping parameters ($t,t',t''$), the Fermi surface indicator  ($r$), the energy difference between the d$_{x^2 - y^2}$ and d$_{3z^2-r2}$ Wannier orbitals ($\Delta E$) ,the in-plane lattice parameter ($a$) and the lattice parameter ratio ($c/a$) are shown. $c$ stands for the out-of-plane lattice parameter.\label{table:big_table}}
  \small
  \begin{tabular}{l|cccc}
  \hline 
  &\multicolumn{4}{|c}{$\alpha$Ba$_2$CuO$_4$F}\\
\cline{2-5}
&AlBa$_2$CuO$_4$F&GaBa$_2$CuO$_4$F&InBa$_2$CuO$_4$F&TlBa$_2$CuO$_4$F\\
\hline \hline
     $t$[eV]&-0.503&-0.493&-0.504&-0.435\\ 
    \hline
     $t'$[eV]&0.094&0.089&0.095&0.126\\ 
     \hline
     $t''$[eV]&-0.065&-0.063&-0.066&-0.070\\ 
     \hline
      $r$&0.316&0.308&0.319&0.451\\ 
     \hline
     $\Delta E$[eV]&1.988&1.921&1.912&2.093\\ 
     \hline
    $U_{\mathrm{bare}}$[eV]&14.364&14.331&14.217&13.422\\ 
        \hline
     $U_{\mathrm{screened}}$[eV]&2.756&2.702&2.315&2.515\\ 
    \hline
     $|U_{\mathrm{screened}}/t|$&5.479&5.481&4.593&5.782\\ 
     \hline
     $a$[\AA]&3.895&3.971&3.916&3.961\\
     \hline
    $c/a$&2.401&2.413&2.470&2.475\\ 
     \hline     \hline
       &\multicolumn{3}{|c|}{$\beta$Ba$_2$CuO$_4$}&\\
\cline{2-5}
&ZnBa$_2$CuO$_4$&CdBa$_2$CuO$_4$&HgBa$_2$CuO$_4$&\multicolumn{1}{|c}{TlBa$_2$CuO$_5$} \\
\hline \hline
     $t$[eV]&-0.454&-0.498&-0.450&-0.574\\ 
    \hline
     $t'$[eV]&0.099&0.094&0.102&0.092\\ 
     \hline
     $t''$[eV]&-0.092&-0.065&-0.095&-0.076\\ 
     \hline
      $r$&0.421&0.319&0.438&0.293\\ 
     \hline
     $\Delta E$[eV]&1.971&1.908&1.799&2.175\\ 
     \hline
    $U_{\mathrm{bare}}$[eV]&12.837&14.322&12.278&13.768\\ 
        \hline
     $U_{\mathrm{screened}}$[eV]&2.795&2.335&2.946&1.708\\ 
    \hline
     $|U_{\mathrm{screened}}/t|$&6.156&4.689&6.547&2.976\\ 
     \hline
     $a$[\AA]&3.965&3.963&3.975&3.826\\
     \hline
    $c/a$&2.461&2.572&2.552&2.603\\ 
     \hline
  \end{tabular}
\end{table*} 

In Table \ref{table:big_table}, we present the evaluated parameters of the compounds. Figure~\ref{fig:band_eight} shows the obtained band structures and Wannier-interpolated bands of the cuprates. From the band structures of TlBa$_2$CuO$_4$F, InBa$_2$CuO$_4$F and AlBa$_2$CuO$_4$F, we can see an extra Fermi surface around the $\Gamma$ point. Some theoretical calculations for Tl$_2$Ba$_2$CuO$_{6+\delta}$ show that this kind of extra Fermi surface disappears around optimal-doping region\cite{Tl2201_band_1,Tl2201_band_2}. Therefore, we assume that an extra Fermi surface around the $\Gamma$ point appearing in each mother compound does not cause a significant effect on the superconductivity.

First, we compare TlBa$_2$CuO$_4$F with TlBa$_2$CuO$_5$. One can see that $r$ for TlBa$_2$CuO$_4$F is large compared to TlBa$_2$CuO$_5$. According to previous studies\cite{nesting2001,nesting2004,sakakibara2012prbpress}, the $t', t''$  values are related to the hybridization between the Cu 3d and 4s orbital. Our calculation shows that energy difference between the Cu:$s$ and Cu:d$_{x^2 - y^2}$ Wannier orbitals is 4.489 eV in TlBa$_2$CuO$_4$F and 5.508 eV in TlBa$_2$CuO$_5$. Therefore, we deduce that this low-level Cu:4s orbital is a reason for the large $t'$ of TlBa$_2$CuO$_4$F. In addition, TlBa$_2$CuO$_5$ have the largest $|t|$ among these cuprates. The amplitude of $t$ is sensitive to $a$-axis lattice parameter so the smallest $a$ of TlBa$_2$CuO$_5$ would be an origin of the largest $|t|$.

Both TlBa$_2$CuO$_5$ and TlBa$_2$CuO$_4$F have similar values of the bare Coulomb interaction, but the former has a much smaller screened Coulomb interaction than the latter
. This fact suggests that screening effect in TlBa$_2$CuO$_5$ is larger than that in TlBa$_2$CuO$_4$F. In terms of the band structures in Fig.~\ref{fig:band_eight}(a)-(c), electronics states of TlBa$_2$CuO$_4$F is closer to HgBa$_2$CuO$_4$ rather than TlBa$_2$CuO$_5$. In particular, dense bands below the Fermi energy do not appear in TlBa$_2$CuO$_4$F unlike TlBa$_2$CuO$_5$, which is considered be one of the origins of smaller screening effect in TlBa$_2$CuO$_4$F.

HgBa$_2$CuO$_4$ exhibits the largest $U_{\mathrm{screened}}$. However, it has the smallest $U_{\mathrm{bare}}$ among the cuprates considered in this research. Thus, we deduce that screening effects in HgBa$_2$CuO$_4$ is small compared to the other cuprates. This strong $U_{\mathrm{screened}}$ is one of the advantages of the high-$T_{\text{c}}$ of HgBa$_2$CuO$_4$. 

$\alpha$Ba$_2$CuO$_4$F($\alpha$=In,Ga,Al) and CdBa$_2$CuO$_4$ have a smaller $r$ than other compounds. This is an important factor for high-temperature superconductivity, because small values of $r$ can enhance spin fluctuations through the nesting of the Fermi surface.  Moreover, GaBa$_2$CuO$_4$F and AlBa$_2$CuO$_4$F have large $U_{\mathrm{screened}}$s. Then We can expect that this combination of small $r$ and large $U_{\mathrm{screened}}$ favors higher-$T_{\text{c}}$
. 

In the Sakakibara criterion, $\Delta E$ is another relevant quantity. The compounds we consider in this research have, however, similar values of $\Delta E$. So, we conclude that it does not make the differences in superconductivity strength among the cuprates considered in this study.

\subsection{Evaluation of superconductivity strength}
\begin{figure}[t]
\centering
    \includegraphics[keepaspectratio, width=90mm]{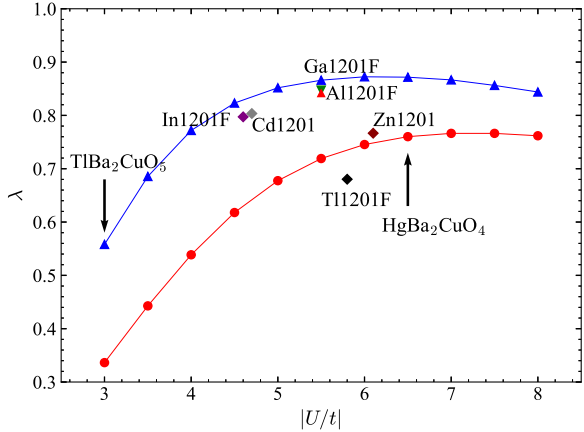}
        \caption{The eigenvalues $\lambda$ of the linearized Eliashberg equation in the single-band Hubbard model. For reference, we plot the red (blue) marks represent the case where hopping parameters derived from DFT are used for Hg1201(Tl1201). The arrows indicate the values of $\lambda$s in Hg1201 and Tl1201 when we use the evaluated $U_{\mathrm{screened}}$ in the constrained-RPA. The red and blue solid lines in the figure are guide for the eye. We write the names of compounds in abbreviated form. \label{fig:FLEX}}
\end{figure}
In Fig. \ref{fig:FLEX}, we show the $\lambda$s obtained by solving the linearized Eliashberg equation within the FLEX approximation. The $U_{\mathrm{screened}}$ obtained from the constrained-RPA calculation is used as the $U$ for the model calculation.

The obtained eigenvalues suggest that TlBa$_2$CuO$_4$F has the potential to exhibit higher-$T_{\text{c}}$ than TlBa$_2$CuO$_5$ owing to the stronger $U$ of  TlBa$_2$CuO$_4$F. Because $\alpha$Ba$_2$CuO$_4$F($\alpha$=In,Ga,Al), and  CdBa$_2$CuO$_4$F have good nesting conditions (small $r$), the values of $\lambda$s are larger than those of Hg1201, although the $U$s are smaller than for Hg1201. $\lambda$ of ZnBa$_2$CuO$_4$F is close to the value of Hg1201 because it has similar $r$ and $U$ to Hg1201. Among these cuprates, those based on Ga and Al stand out conspicuously.

\subsection{Stability of fluorine-doped materials}
\begin{table}[ht]
  \caption{Relative total energy of F substitution at each non-equivalent oxygen site in TlBa$_2$CuO$_5$. Fluorine substitution at the buffer layer site(site 1) shows the lowest total energy, so this is set to 0.\label{table:tot_energy_site}}
  \small
  \begin{tabular}{|l|c|}
\hline
&Relative energy~[eV]\\
\hline \hline
Buffer layer (Site 1)& 0.0\\ 
Apical (Site 2)& 0.970 \\ 
CuO$_2$ layer (Site 3)&  0.876 \\ 
     \hline
  \end{tabular}
\end{table} 

In this section, we discuss stability of the fluorine-doped materials. When replacing one oxygen with fluorine in TBa$_2$CuO$_5$, there are three possible sites, namely the oxygen in the buffer layer (Site 1), the apical oxygen site (Site 2), and the oxygen in the CuO$_2$ layer (Site 3), as shown in Fig.\ref{fig:str}. In Table \ref{table:tot_energy_site}, we show the relative total energies for F substitution into three different oxygen sites. One can see that F substitution into the buffer layer shows the lowest energy.

Here, we consider the formation energy of fluorine doped into the buffer layer. The formation energy is defined as
\begin{equation}
E_{\rm{Form}}=E[\mathrm{TlBa_2CuO_4F}]+\frac{1}{2}E[\mathrm{O}_{2}]-(E[\mathrm{TlBa_2CuO_5}]+\frac{1}{2}E[\mathrm{F}_{2}]),
\end{equation}
where $E[:]$ is the total energy\cite{cuprates_formationenergy}. The obtained value of the formation energy is $E_{\rm{Form}} = -1.630$ eV\cite{comment}, meaning that fluorine-doped TlBa$_2$CuO$_4$F is more stable than TlBa$_2$CuO$_5$.
Thus, we conclude that fluorine doping into the buffer layer is feasible.

\section{Summary and discussion}
In this research, we analyzed various cuprates, including the structures that have not been synthesized. Al and Ga compounds were found to show a high potential for high-$T_{\text{c}}$ superconductivity. In and Cd compounds also possessed good parameters for high-$T_{\text{c}}$. According to our model calculations, these cuprates have the potential to exhibit superconductivity at higher temperatures than Hg1201. If our proposed crystal structures are experimentally synthesized, they may break the world record for $T_{\text{c}}$ in single-layer cuprates. Moreover, comparison of  TlBa$_2$CuO$_5$ and TlBa$_2$CuO$_4$F shows that $U$ and $r$ are controllable by changing the components and dopants in the buffer layer. 

There are some reports that show the crystal phases related to the compounds considered in this research although there remains a problem in making the perfect crystal. For example, as for In-based cuprates, the relevant crystal phases have been synthesized\cite{In1201_2017}. Cd-based cuprates are synthesized and they show high-$T_{\text{c}}$ superconductivity\cite{cd2,cd3,cd4}. Multi-layered cuprates including Al or Ga are reported\cite{Albased,Gabased}.

The determination of materials parameters is also
useful for considering additional effects of multi-layer systems\cite{Nishiguchi2013,Nishiguchi2017,Nishiguchi2018}.

Suppression of $T_{\text{c}}$ by disorder effects has been reported in cuprates\cite{Potential_eisaki}. One advantage of controlling parameters by changing the components and dopants in the buffer layer is that those components and dopants do not directly affect the CuO$_2$ layer due to the distance between the buffer layer and the CuO$_2$ layer. 

\begin{acknowledgment}
The calculations were performed in the computer centers of Kyushu University and ISSP, University of Tokyo.
\end{acknowledgment}

\bibliographystyle{jpsj}
\bibliography{newstr}

\begin{thebibliography}{10}

\bibitem{Hg1201-first}
S.~Putilin, E.~Antipov, O.~Chmaissem, and M.~Marezio: Nature {\bfseries 362}
  (1993) 226.

\bibitem{Sakakibara2010}
H.~Sakakibara, H.~Usui, K.~Kuroki, R.~Arita, and H.~Aoki: Phys. Rev. Lett.
  {\bfseries 105} (2010) 2.

\bibitem{sakakibara2012}
H.~Sakakibara, H.~Usui, K.~Kuroki, R.~Arita, and H.~Aoki: Phys. Rev. B
  {\bfseries 85} (2012) 064501.

\bibitem{Sakakibara2014}
H.~Sakakibara, K.~Suzuki, H.~Usui, S.~Miyao, I.~Maruyama, K.~Kusakabe,
  R.~Arita, H.~Aoki, and K.~Kuroki: Phys. Rev. B {\bfseries 89} (2014) 6.

\bibitem{cRPAteranishi}
S.~Teranishi, K.~Nishiguchi, and K.~Kusakabe: Journal of the Physical Society
  of Japan {\bfseries 87} (2018) 114701.

\bibitem{wannier_ent}
I.~Souza, N.~Marzari, and D.~Vanderbilt: Phys. Rev. B {\bfseries 65} (2001)
  035109.

\bibitem{wannier_loc}
N.~Marzari and D.~Vanderbilt: Phys. Rev. B {\bfseries 56} (1997) 12847.

\bibitem{wannier90}
A.~A. Mostofi, J.~R. Yates, G.~Pizzi, Y.-S. Lee, I.~Souza, D.~Vanderbilt, and
  N.~Marzari: Computer Physics Communications {\bfseries 185} (2014) 2309 .

\bibitem{cRPA}
F.~Aryasetiawan, M.~Imada, A.~Georges, G.~Kotliar, S.~Biermann, and A.~I.
  Lichtenstein: Phys. Rev. B {\bfseries 70} (2004) 195104.

\bibitem{cRPA_cuprates_Aryasetiawan}
F.~Nilsson, K.~Karlsson, and F.~Aryasetiawan: Phys. Rev. B {\bfseries 99}
  (2019) 075135.

\bibitem{FLEX_1}
N.~E. Bickers, D.~J. Scalapino, and S.~R. White: Phys. Rev. Lett. {\bfseries
  62} (1989) 961.

\bibitem{FLEX_2}
N.~Bickers and D.~Scalapino: Annals of Physics {\bfseries 193} (1989) 206 .

\bibitem{Hg1201_F_Tc}
A.~M. Abakumov, V.~L. Aksenov, V.~A. Alyoshin, E.~V. Antipov, A.~M. Balagurov,
  D.~A. Mikhailova, S.~N. Putilin, and M.~G. Rozova: Phys. Rev. Lett.
  {\bfseries 80} (1998) 385.

\bibitem{QE-2009}
P.~Giannozzi, S.~Baroni, N.~Bonini, M.~Calandra, R.~Car, C.~Cavazzoni,
  D.~Ceresoli, G.~L. Chiarotti, M.~Cococcioni, I.~Dabo, A.~{Dal Corso},
  S.~de~Gironcoli, S.~Fabris, G.~Fratesi, R.~Gebauer, U.~Gerstmann,
  C.~Gougoussis, A.~Kokalj, M.~Lazzeri, L.~Martin-Samos, N.~Marzari, F.~Mauri,
  R.~Mazzarello, S.~Paolini, A.~Pasquarello, L.~Paulatto, C.~Sbraccia,
  S.~Scandolo, G.~Sclauzero, A.~P. Seitsonen, A.~Smogunov, P.~Umari, and R.~M.
  Wentzcovitch: Journal of Physics: Condensed Matter {\bfseries 21} (2009)
  395502 (19pp).

\bibitem{QE-2017}
P.~Giannozzi, O.~Andreussi, T.~Brumme, O.~Bunau, M.~B. Nardelli, M.~Calandra,
  R.~Car, C.~Cavazzoni, D.~Ceresoli, M.~Cococcioni, N.~Colonna, I.~Carnimeo,
  A.~D. Corso, S.~de~Gironcoli, P.~Delugas, R.~A.~D. Jr, A.~Ferretti,
  A.~Floris, G.~Fratesi, G.~Fugallo, R.~Gebauer, U.~Gerstmann, F.~Giustino,
  T.~Gorni, J.~Jia, M.~Kawamura, H.-Y. Ko, A.~Kokalj,
  E.~K\"{u}\c{c}\"{u}kbenli, M.~Lazzeri, M.~Marsili, N.~Marzari, F.~Mauri,
  N.~L. Nguyen, H.-V. Nguyen, A.~O. de-la Roza, L.~Paulatto, S.~Ponc\'e,
  D.~Rocca, R.~Sabatini, B.~Santra, M.~Schlipf, A.~P. Seitsonen, A.~Smogunov,
  I.~Timrov, T.~Thonhauser, P.~Umari, N.~Vast, X.~Wu, and S.~Baroni: Journal of
  Physics: Condensed Matter {\bfseries 29} (2017) 465901.

\bibitem{respack_1}
K.~Nakamura, Y.~Nohara, Y.~Yoshimoto, and Y.~Nomura: Phys. Rev. B {\bfseries
  93} (2016) 085124.

\bibitem{respack_2}
K.~Nakamura, Y.~Yoshimoto, T.~Kosugi, R.~Arita, and M.~Imada: J. Phys. Soc. Jpn
  {\bfseries 78} (2009) 083710.

\bibitem{respack_3}
K.~Nakamura, R.~Arita, and M.~Imada: J. Phys. Soc. Jpn {\bfseries 77} (2008)
  093711.

\bibitem{respack_4}
Y.~Nohara, S.~Yamamoto, and T.~Fujiwara: Phys. Rev. B {\bfseries 79} (2009)
  195110.

\bibitem{respack_5}
T.~Fujiwara, S.~Yamamoto, and Y.~Ishii: J. Phys. Soc. Jpn {\bfseries 72} (2003)
  777.

\bibitem{Tl2201_band_1}
S.~Sahrakorpi, H.~Lin, R.~Markiewicz, and A.~Bansil: Physica C:
  Superconductivity and its applications {\bfseries 460} (2007) 428.

\bibitem{Tl2201_band_2}
D.~Peets, J.~Mottershead, B.~Wu, I.~Elfimov, R.~Liang, W.~Hardy, D.~Bonn,
  M.~Raudsepp, N.~Ingle, and A.~Damascelli: New Journal of Physics {\bfseries
  9} (2007) 28.

\bibitem{nesting2001}
E.~Pavarini, I.~Dasgupta, T.~Saha-Dasgupta, O.~Jepsen, and O.~K. Andersen:
  Phys. Rev. Lett. {\bfseries 87} (2001) 047003.

\bibitem{nesting2004}
K.~Tanaka, T.~Yoshida, A.~Fujimori, D.~H. Lu, Z.-X. Shen, X.-J. Zhou,
  H.~Eisaki, Z.~Hussain, S.~Uchida, Y.~Aiura, K.~Ono, T.~Sugaya, T.~Mizuno, and
  I.~Terasaki: Phys. Rev. B {\bfseries 70} (2004) 092503.

\bibitem{sakakibara2012prbpress}
H.~Sakakibara, K.~Suzuki, H.~Usui, K.~Kuroki, R.~Arita, D.~J. Scalapino, and
  H.~Aoki: Phys. Rev. B {\bfseries 86} (2012) 134520.

\bibitem{cuprates_formationenergy}
J.~Gazquez, R.~Guzman, R.~Mishra, E.~Bartolomé, J.~Salafranca, C.~Magén,
  M.~Varela, M.~Coll, A.~Palau, S.~M. Valvidares, P.~Gargiani, E.~Pellegrin,
  J.~Herrero-Martin, S.~J. Pennycook, S.~T. Pantelides, T.~Puig, and
  X.~Obradors: Advanced Science {\bfseries 3} (2016) 1500295.

\bibitem{comment}
For the energy calculation of molecules, we carried out several
  calculations(LSDA, B3LYP and PBE0) and the obtained values of the formation
  energies are all negative. The value of formation energy listed is the one
  that is the closest to 0. We take the the value of 20 \AA$~$for the vacuum
  gap.

\bibitem{In1201_2017}
Y.~Watanabe, N.~Komiyama, Y.~Shimabukuro, M.~Satoh, and S.~Kambe: Transactions
  of the Materials Research Society of Japan {\bfseries 42} (2017) 159.

\bibitem{cd2}
N.~Balchev, V.~Lovchinov, E.~Gattef, A.~Staneva, K.~Konstantinov, and J.~Pirov:
  Journal of Superconductivity {\bfseries 8} (1995) 329.

\bibitem{cd3}
N.~Balchev, V.~Lovchinov, E.~Gattef, and A.~Staneva: Journal of
  superconductivity {\bfseries 8} (1995) 333.

\bibitem{cd4}
R.~Mariychuk, P.~Popovich, V.~Bunda, S.~Meszaros, and E.~Semrad: Bulgarian
  Journal of Physics {\bfseries 27} (2000) 33.

\bibitem{Albased}
M.~Isobe, T.~Kawashima, K.~Kosuda, Y.~Matsui, and E.~Takayama-Muromachi:
  Physica C: Superconductivity {\bfseries 234} (1994) 120 .

\bibitem{Gabased}
E.~Takayama-Muromachi and M.~Isobe: Japanese Journal of Applied Physics
  {\bfseries 33} (1994) L1399.

\bibitem{Nishiguchi2013}
K.~Nishiguchi, K.~Kuroki, R.~Arita, T.~Oka, and H.~Aoki: Phys. Rev. B
  {\bfseries 88} (2013) 014509.

\bibitem{Nishiguchi2017}
K.~Nishiguchi, S.~Teranishi, and K.~Kusakabe: J. Phys. Soc. Jpn {\bfseries 86}
  (2017) 084707.

\bibitem{Nishiguchi2018}
K.~Nishiguchi, S.~Teranishi, K.~Kusakabe, and H.~Aoki: Phys. Rev. B {\bfseries
  98} (2018) 174508.

\bibitem{Potential_eisaki}
H.~Eisaki, N.~Kaneko, D.~L. Feng, A.~Damascelli, P.~K. Mang, K.~M. Shen, Z.-X.
  Shen, and M.~Greven: Phys. Rev. B {\bfseries 69} (2004) 064512.

\end{thebibliography}

\end{document}